# A study on the thermal conductivity of compacted bentonites

**Anh-Minh TANG, Yu-Jun CUI, Trung-Tinh LE**
*Ecole des Ponts ParisTech, UR Navier/CERMES, 6 et 8, av. Blaise Pascal, Cité Descartes, Champs-sur-Marne, 77455 Marne La Vallée CEDEX 2, France.*

*Abstract*

Thermal conductivity of compacted bentonite is one of the most important properties in the design of high-level radioactive waste repositories where this material is proposed for use as a buffer. In the work described here, a thermal probe based on the hot wire method was used to measure the thermal conductivity of compacted bentonite specimens. The experimental results were analyzed to observe the effects of various factors (i.e. dry density, water content, hysteresis, degree of saturation and volumetric fraction of soil constituents) on the thermal conductivity. A linear correlation was proposed to predict the thermal conductivity of compacted bentonite based on experimentally observed relationship between the volumetric fraction of air and the thermal conductivity. The relevance of this correlation was finally analyzed together with others existing methods using experimental data on several compacted bentonites.

## *1. Introduction*

Compacted bentonite is often considered as a possible buffer material for high-level radioactive waste disposal. Its thermal conductivity is one of the key properties for the design of such disposal system (JNC, 2000). Several works have been done previously to study the thermal conductivity of compacted bentonites. Measured data can be found in the works of Villar (2000) on Febex bentonite, Ould-Lahoucine et al. (2002) on Kunigel bentonite, Madsen (1998) on MX80 bentonite, Coulon et al. (1987) on several smectite-based clays. These measurements show that the thermal conductivity of compacted bentonites depends on the dry density, the water content, and the mineralogical composition. These parameters, which are in general easily measurable, have been often used in order to predict the thermal conductivity of soil (Johansen, 1975; De Vries, 1963, Ould-Lahoucine et al., 2002; among others).

In order to verify the relevance of various prediction methods, a study on the thermal conductivity properties of compacted MX80 bentonite was performed. A commercial thermal analyzer that conforms to the ASTM Standard was used to measure the thermal conductivity of MX80 bentonite compacted at various dry densities and water contents. The mineralogical composition of the MX80 bentonite studied in the present work is different from that presented by Madsen (1998). Therefore, particular attention has been paid to the effect of mineralogical composition. The effect of various factors was analyzed. This analysis confirmed the observation of Ochsner et al. (2001) on various soils: the soil thermal conductivity is strongly correlated with the volume fraction of air. A linear correlation was then proposed to predict the thermal conductivity of compacted bentonites. The relevance of this correlation was finally analyzed in comparison with existing approaches.

## 2. Experimental methodology

### 2.1. Material studied

MX80 bentonite, from Wyoming, has been considered as a potential reference buffer material in Sweden, Switzerland and others countries (Pusch and Yong, 2003). In the study described here, the MX80 bentonite was purchased from CETCO Europe Ltd in 2004. Identification parameters of this soil are presented in Table 1. This table presents equally the properties of two others bentonites (Febex and Kunigel) that are proposed as potential buffer material in Spain and in Japan as well as the properties of MX80 bentonite presented by Madsen (1998). It appears that the proportion of smectite is dominant in the bentonite. A significant difference in the proportion of quartz has been reported for the two MX80 bentonites: Madsen (1998) studied the MX80 bentonite provided by Bentonit International GmbH, Duisburg, Germany and found 15 % of quartz; whereas the MX80 bentonite studied in the present work (provided by CETCO Europe Ldt) contained only 3 % of quartz. With their high plasticity index ($Ip$) and specific surface area ($S$), these soils are classified as very highly plastic clays.

### 2.2. Experiments

To analyze the effect of different factors on the thermal conductivity of compacted bentonite specimens, several procedures of preparation were applied. In total, 16 specimens were prepared by compacting the bentonite into a mould using a standard press. Table 2 presents the preparing procedures of all specimens. Prior to compaction, the provided bentonite was firstly sieved at 2 mm. The sieved bentonite was then placed at constant temperature (20°C) with controlled relative humidity ($RH$). The relative humidity was controlled by air circulation between the vessel and a bottle containing a saturated saline solution. This method using vapour equilibrium technique was described by Delage et al. (1998) and Tang and Cui (2005). The samples were weighted every 3 days until no further weight change occurred. The prepared bentonite was then compacted statically in a mould at a constant displacement rate of 0.1 mm/min, using a 50 kN digitally controlled press. During compaction, the applied pressure was measured using a force transducer and the press was stopped when the dry density of bentonite reached the target value. As the water content of the bentonite varied from 9 to 18% and the dry density varied from 1.4 to 1.8 Mg/m$^3$, the maximal compaction pressure applied varied from 6 to 40 MPa. The dimension of the compacted specimens was 50 mm in diameter and 70 mm high. The duration of compaction was about 2 hours.

After measuring the thermal conductivity, the specimens of series S2 and S3 were put in a hermetically-sealed box having 44 % $RH$. This operation caused a decrease of water content in the specimens. Five months were required for their mass stabilization. The thermal conductivity of these specimens was then measured again prior to oven-drying at 105 °C to determine the final gravimetric water content. These series are presented as series S2b and S3b in Table 2.

### 2.3. Thermal conductivity measurement method

The instrument used to measure the thermal conductivity is a commercial thermal properties analyzer, KD2 (Decagon Devices Inc.). Its operating concept is based on the hot wire method where the thermal conductivity is calculated by monitoring the heat dissipation from a linear heat source at a given voltage (ASTM D 5334-00, 2000). The probe consists of a heating wire (60 mm long and 1.28 mm in diameter) and a thermistor in the middle of the wire. During the measurement, the controller firstly heats the probe for 30 s and then calculates the thermal characteristics. After Decagon Devices Inc. (2006), this device conforms to ASTM Standards, ASTM D5334-00 (2000).

After compaction, a hole of 1.3 mm in diameter and 6 mm deep was drilled into the middle of the specimen. To provide better thermal contact between the sample and the probe, the probe was coated within a thin layer of thermal grease. The set-up for the thermal conductivity measurement and the specifications of the KD2 instrument are presented in Figure 1.

## 3. Experimental results

### 3.1. Effect of dry density, water content and mineralogy

The results on thermal conductivity of Series S1, S2, S3 are plotted in Figure 2 versus dry density. The results of Kahr and Müller – Vonmoos (1982), cited by Madsen (1998), are also presented in Figure 2. Kahr and Müller – Vonmoos (1982) measured the thermal conductivity of MX80 at two water contents: 7 % (Series KM1 in Figure 2) and 14 % (Series KM2 in Figure 2).

The thermal conductivity ($K$) was proportional to the dry density ($\rho_d$) for a given series: the higher the dry density the higher the thermal conductivity. On the other hand, the effect of water content ($w$) was evident: at the same dry density, the higher the water content the higher the thermal conductivity. In fact, the $K - \rho_d$ plot of series S3 ($w$ = 17.9 %) was in a higher position than that of series S2 ($w$ = 11.7 %); the latter was in a higher position than that of series S1 ($w$ = 9.0 %). Thus, it is clear that the thermal conductivity increased with dry density or water content increase.

Although the effect of dry density and water content was evident from the results of all the test series S1, S2, S3, comparison between the results obtained from the present work and that by Kahr and Müller – Vonmoos (1982) showed a difference between these two bentonites. The $K - \rho_d$ plot of series KM1 ($w$ = 7%) was in a higher position than that of series S1 having higher water content (9.0 %); likewise the $K - \rho_d$ plot of series KM2 ($w$ = 14 %) was in a higher position than that of series S3 having higher water content (17.9 %). Thus, the thermal conductivity of the bentonite studied by Kahr and Müller – Vonmoos (1982) was higher than that studied in the present work at the same values of dry density and water content.

As Madsen (1998) did not present any details about the experimental technique used in the work of Kahr and Müller – Vonmoos (1982), no comment about the effect of the experimental technique on the results can be made to explain the observed difference. However, the difference in the mineralogical composition of these two MX80 bentonites studied in these works may explain the difference of the thermal conductivity. As shown in Table 1, The MX80 presented by Madsen (1998) contained 15 % of quartz, while that in the present work contained only 3 % of quartz. As the thermal conductivity of quartz (7.7 W/mK) is much higher than that of other minerals (2.0 W/mK), (Johansen, 1975), the proportion of quartz may significantly affect the thermal conductivity. This effect was discussed by Farouki (1986) and taken into account in the model of Johansen (1975).

**3.2. Hysteresis effects**

In Figure 3 the results of series S1, S2, S3 are presented together with that of series S2b, S3b. The $K - \rho_d$ plot of series S2 ($w = 11.7$ %) was identical to that of series S3b having a lower water content ($w = 10.2$%). In addition, the $K - \rho_d$ plot of series S2b ($w = 9.1$%) was clearly in a higher position than that of series S1 having similar water content ($w = 9.0$ %). The thermal conductivity of dried specimens (series S2b and S3b) was higher than the compacted specimens (series S1, S2, S3) having the same dry density and water content.

Farouki (1986) also noted that the thermal conductivity of soil at the same dry density and water content depends on whether this water content was reached by wetting or drying: when it was produced by drying, the thermal conductivity was much higher. Delage et al. (2006) performed some observations on the microstructure of compacted MX80 bentonite and observed that compaction corresponds to a suction decrease. Thus, it can be considered that the specimens in the series S1, S2 were produced by wetting, while the water content in the series S2b and S3b were reached by drying. The results obtained are then in agreement with that noted by Farouki (1986).

**3.3. Effect of components**

In Figure 4, the thermal conductivity is presented versus: (a) degree of saturation, $S_r$; (b) solids volume fraction, $V_s/V$; (c) water volume fraction, $V_w/V$; (d) air volume fraction, $V_a/V$. There were no obvious relationships observed between $K$ and $S_r$, $V_s/V$, or $V_w/V$. However, the thermal conductivity values of the specimens having the same $V_a/V$ were identical; when $V_a/V$ increased, $K$ decreased in a linear fashion. This trend was also noted by Ochsner et al. (2001) for different soils. The thermal conductivity of the solids and that of water ($K_s = 2.0$ W/mK for the minerals and $K_w = 0.57$ W/mK for water) are in the same order of magnitude, and are much higher than the thermal conductivity of air ($K_a = 0.025$ W/mK), as a result, $K$ is strongly correlated with $V_a/V$. For a given $V_a/V$, increasing $V_w/V$ decreases $V_s/V$, and therefore reduces the thermal conductivity. The increase of $V_w/V$ may also improve thermal contacts between the solid particles and thus increases the thermal conductivity. Therefore, the combination of these two opposing phenomena reduced the influence of the water content on the change of thermal conductivity.

## *4. Calculating the thermal conductivity*

**4.1. Johansen's method**

According to Farouki (1986), the method developed by Johansen (1975) was applicable for unfrozen fine-grained soils at $S_r$ higher than 20 %. The thermal conductivity ($K$) was expressed as

$$K = (K_{sat} - K_{dry})K_e + K_{dry}$$

where $K_{sat}$ and $K_{dry}$ are respectively the thermal conductivity in saturated and dry states at a same dry density, $K_e$ is a function representing the influence of $S_r$ on $K$: $K_e = 1.0 + \log_{10} S_r$. For saturated unfrozen soils: $K_{sat} = K_s^{(1-n)} K_w^n$, where $n$ is the porosity and $K_w$ is the thermal conductivity of water, $K_w = 0.57$ W/mK. The thermal conductivity of the solids $K_s$ is calculated by using the equation: $K_s = K_q^q K_0^{1-q}$, where $K_q$ is the thermal conductivity of quartz ($K_q = 7.7$ W/mK), $K_0$ is the thermal conductivity of other minerals ($K_0 = 2.0$ W/mK) and $q$ is the quartz volume fraction. For dry soils: $K_{dry} = \dfrac{0.135\rho_d + 64.7}{\rho_s - 0.947\rho_d} (W/mK)$, where the dry unit weight, $\rho_d$, and the unit weight of the solids, $\rho_s$, are expressed in kg/m$^3$.

To evaluate this method, the calculated thermal conductivity values were compared with the measured values taken from the present work and from the literature: MX80 (Kahr and Müller-Vonmoos, 1982); Febex (Villar, 2000); Kunigel (Ould-Lahoucine et al., 2002). Only the specimens having $S_r > 20$ % have been chosen. The identification parameters of these bentonites are presented in Table 1. The thermal conductivity of the solids was calculated using the proportion of quartz presented in Table 1: $K_s = 2.08$ W/mK for the MX80 of the present work; $K_s = 2.45$ W/mK (for the MX80 in the work of Kahr and Müller-Vonmoos, 1982), $K_s = 2.05$ W/mK for the Febex bentonite; and $K_s = 3.12$ W/mK for the Kunigel bentonite.

Figure 5 presents the comparison between the calculation by Johansen's method and the experimental data. In general, this method overestimated the thermal conductivity of compacted bentonites. For a better quantification of the comparison, two factors were calculated: mean value of $R$ ($a = \dfrac{1}{N}\sum_{i=1}^{N} R_i$) and root mean square error of $R$ ($b = \sqrt{\dfrac{1}{N}\sum_{i=1}^{N}(R_i - 1)^2}$) where $R$ was defined as the ratio of the predicted value to the experimental value. As a result, the prediction of the method can be considered good if $a \cong 1$ and $b \cong 0$. Ould-Lahoucine et al. (2002) also used these two factors for assessing model accuracy. As indicated by Figure 5, the method of Johansen (1975) gave the best prediction with the Febex bentonite ($a = 1.11$, $b = 0.19$). For the other bentonites, the model overestimated the thermal conductivity by more than 20 % (i.e. $a > 1.2$).

Ould-Lahoucine et al. (2002) compared the experimental data on several compacted bentonites (MX80 and Kunigel) with the correlation of Knutsson (1983) which is similar to the Johansen's model (1975): the only difference is that Knutsson (1983) have chosen the thermal conductivity of the solids, $K_s$, equal to 2.0 W/mK. With this value, Ould-Lahoucine et al. (2002) obtained better prediction than Johansen (1975) who took a = 0.907 and b = 0.152 for the Kunigel and the MX80 (see Figure 5).

## 4.2. De Vries's model

De Vries (1963) proposed a method that uses the weighted average of the thermal conductivity value of each soil constituent. For unsaturated soils, solids particles and air are considered to be two components immersed in the continuous media: water. This assumption applies when the volumetric water fraction is above a certain minimum limit so that water can be considered as continuous. For fine-grained soils, this limit is $V_w/V = 0.05$ to $0.10$. The thermal conductivity of such a soil is expressed as:

$$K = \frac{(V_w/V)K_w + F_a(V_a/V)K_a + F_s(V_s/V)K_s}{(V_w/V) + F_a(V_a/V) + F_s(V_s/V)}$$

where $F_s, F_a$ are the weighing factors depending on the shape and the orientation of soil particles and air-pores, respectively. According to Farouki (1986),

$$F_s = \frac{1}{3}\left\{\frac{2}{1+\left(\frac{K_s}{K_w}-1\right)0.125} + \frac{1}{1+\left(\frac{K_s}{K_w}-1\right)0.75}\right\}$$

$$F_a = \frac{1}{3}\left\{\frac{2}{1+\left(\frac{K_a}{K_w}-1\right)g_a} + \frac{1}{1+\left(\frac{K_a}{K_w}-1\right)g_c}\right\}$$

where $g_a, g_c$ are called shape factors: $g_a = 0.333 - \frac{V_a/V}{n}(0.333 - 0.035)$ and $g_c = 1 - 2g_a$.

Calculations with the Johansen's method proved that the choice of the thermal conductivity of the solids was important in the prediction. When applying the De Vries's model, Ochsner et al. (2001) chose the $K_s$ for each soil so that the modelled and the measured values were identical for the most saturated sample of that soil. With this choice, Ochsner et al. (2001) obtained a good agreement between the prediction and the measurement of four medium-textured soils covering large ranges of volume fraction of components. By applying this method, the following value of $K_s$ was chosen: $K_s = 1.5$ W/mK for the MX80 in the present work; $K_s = 1.9$ W/mK for the other bentonites (MX80 from Kahr and Müller-Vonmoos, 1982; Febex from Villar, 2000; Kunigel from JNC, 2000; see Table 1).

Figure 6 presents the results calculated using the De Vries's model versus the data (only the specimens having $V_w/V > 0.10$ were chosen for the comparison). The De Vries's model gave a better prediction than the Johansen's model because it overestimated only 10 % the thermal conductivity of MX80 in the present work ($a = 1.11$), of Febex ($a = 1.09$) and of MX80 in the work of Kahr and Müller-Vonmoos (1982) ($a = 1.10$). It gave a good prediction of the data on Kunigel ($a = 1.02$, $b = 0.08$).

**4.3. Sakashita and Kumada's model**

Sakashita and Kumada (1998) proposed a heat transfer model that accounts for the microstructure of compacted bentonites. The intraparticle micropores and the interparticles macropores are assumed to be of rectangular parallelepiped shape dispersed in a continuous solid phase. Ould-Lahoucine et al. (2002) determined the unknown constants included in the model using experimental data and revised the model to the following semi-theoretical form:

$$K = K_{dry}\left\{1 + \left[(9.750n - 0.706)S_r\right]^{0.285n+0.731}\right\}$$

where the thermal conductivity of dry bentonite is calculated by
$$K_{dry} = 0.0497 + 0.222(1-n) + 0.968(1-n)^3.$$
The results calculated by the model of Sakashita and Kumada (1998) using the parameters determined by Ould-Lahoucine et al. (2002) are presented in Figure 7 versus the measured values. This correlation overestimated the thermal conductivity of MX80 in the present work and the Febex bentonite, but gave a very good prediction for the MX80 bentonite in the work of Kahr and Müller-Vonmoos (1982) and for the Kunigel bentonite (Figure 7).

## 4.4. Formulation of a linear correlation

The results presented in Figure 4d showed that the thermal conductivity of compacted MX80 bentonite can be satisfactorily described as a decreasing linear function of the air-pore volume fraction. Based on this observation, a linear relationship can be formulated as follows
$$K = \alpha(V_a/V) + K_{sat}$$
where $\alpha$ is the slope of $K - V_a/V$ plot; $K_{sat}$ is the thermal conductivity at saturated state which corresponds to the intersection of $K - V_a/V$ plot with $K$ axis. Hypothetically, in a pure air media: $K_a = \alpha + K_{sat}$. Because the thermal conductivity of air is close to zero, the relationship $\alpha \approx -K_{sat}$ can be hypothesized.

In order to calibrate the two parameters for each bentonite ($\alpha$ and $K_{sat}$), two measurements corresponding to the maximum and the minimum values of $V_a/V$ were chosen to fit the model. The calibrated parameters are presented in Table 3. The relationship $\alpha \approx -K_{sat}$ was not really verified, which probably indicates that the observed linear relationship between $K$ and $V_a/V$ is not valid for $V_a/V = 1$.

Figure 8 presents the thermal conductivity calculated using the linear correlation versus the measured values. It can be observed that the linear correlation satisfactorily predicted the measured values of all the specimens with an error less than 20 %. Concerning the mean value of $R$ or $a$, this correlation overestimated 3 % to 7 % the thermal conductivity of these four bentonites ($a = 1.03 - 1.07$) with a dispersion, $b$, less than 15 % ($b = 0.05 - 0.13$).

## 4.5. Discussions

The method proposed by Johansen (1975) can give a good prediction with unfrozen fine soils after Farouki (1986). Nevertheless, Figure 5 shows that this model overestimated the thermal conductivity of compacted bentonites ($a = 1.11 - 1.30; b = 0.19 - 0.31$). In fact, in this method, the thermal conductivity of the solids ($K_s$) was calculated as a function of the proportion of quartz. After Ould-Lahoucine et al. (2002), by imposing the proportion of quartz equal to zero in the calculation of $K_s$, i.e. $K_s = 2.0$ W/mK, this method can give a better prediction for the Kunigel and the MX80 bentonites: $a = 0.907; b = 0.152$.

After Ochsner et al. (2001), the model proposed by De Vries (1963) can give a good prediction if the thermal conductivity of the solids ($K_s$) is calibrated to fit the measured value of the most saturated sample. Figure 6 shows the prediction of this method on compacted bentonites. The prediction of the De Vries's method ($a = 1.02 – 1.11$; $b = 0.08 – 0.18$) was better than that of Johansen's method. The values of $K_s$ adjusted to fit the measured thermal conductivity of the most saturated samples ($K_s = 1.5$ W/mK for the MX80 in the present work; $K_s = 1.9$ W/mK for others bentonites) were close to the value proposed by Knutsson (1983) ($K_s = 2.0$ W/mK).

The correlation used by Ould-Lahoucine et al. (2002) based on the model of Sakashita and Kumada (1998) gave a good prediction for the Kunigel bentonite but a relatively poor prediction for the other bentonites. In fact, Ould-Lahoucine et al. (2002) fitted the model of Sakashita and Kumada (1998) to the experimental data on the Kunigel bentonite to establish this correlation. That explains the performance of this correlation when predicting the thermal conductivity of Kunigel bentonite.

The linear correlation proposed in the present work required at least two measured values to adjust the parameters. The parameters gave a good prediction for all bentonites ($a = 1.03 – 1.07$; $b = 0.05 – 0.13$) (see Figure 8). Note however that the proposed linear correlation requires two calibration points; the De Vries's method requires one and the Johansen's method don't need the calibration; that may partly explain the better performance of the proposed linear correlation.

In the linear correlation, $K_{sat}$ mainly depends on the mineralogical composition and porosity. Therefore it could be determined indirectly using parameters such as quartz fraction, porosity etc. (see Johansen's method). On the contrary, the parameter $\alpha$ depends on soil fabric and is thus difficult to be predicted from other easily measurable soil properties.

## 5. Conclusion

The thermal conductivity of compacted MX80 bentonite was measured using the heat wire method. The effect of the mineralogical composition was evident: the MX80 bentonite studied here contained a lower fraction of quartz than that studied by Madsen (1998) and had lower thermal conductivity. Water content, dry density, hysteresis, degree of saturation and volumetric fraction of constituents are also of influence. A good correlation between the volume fraction of air and thermal conductivity was observed. A linear correlation was proposed to predict the thermal conductivity of compacted bentonites.

## 6. Acknowledgement

The authors are grateful to Ecole Nationale des Ponts et Chaussées (ENPC) and French Electricity Company (EDF) for their financial support.

## Figure and Table captions

| Clay | MX80 (Present work) | MX80 (Madsen, 1998) | Febex (Villar, 2000) | Kunigel (JNC, 2000) |
|---|---|---|---|---|
| Smectite (%) | 92 | 76 | 92 | 46-49 |
| Quartz (%) | 3 | 15 | 2 | 29-38 |
| $w_L$ (%) | 520 | - | 102 | 415 |
| $w_P$ (%) | 42 | - | 53 | 32 |
| Ip | 478 | - | 49 | 383 |
| $\rho_s$ (Mg/m$^3$) | 2.76 | 2.76 | 2.70 | 2.79 |
| S (m$^2$/g) | - | 562 | 725 | 687 |

Table 1. Identification parameters.

| Series | Number | Water content (%) | Preparation method |
|---|---|---|---|
| S1 | 7 | 9.0±0.1 | Dried at 44 % RH |
| S2 | 5 | 11.7±0.1 | Dried at 54 % RH |
| S3 | 4 | 17.9±0.1 | Dried at 76 % RH |
| S2b | 5 | 9.1±0.1 | Series S2 dried at 44 % RH |
| S3b | 4 | 10.2±0.1 | Series S3 dried at 44 % RH |

Table 2. Experimental program.

| Soil | $K_{sat}$ (W/mK) | $\alpha$ (W/mK) |
|---|---|---|
| Present work | 1.10 | -1.79 |
| Febex | 1.30 | -2.29 |
| MX80 | 1.38 | -2.35 |
| Kunigel | 1.39 | -2.36 |

Table 3. Parameters used in the linear correlation.

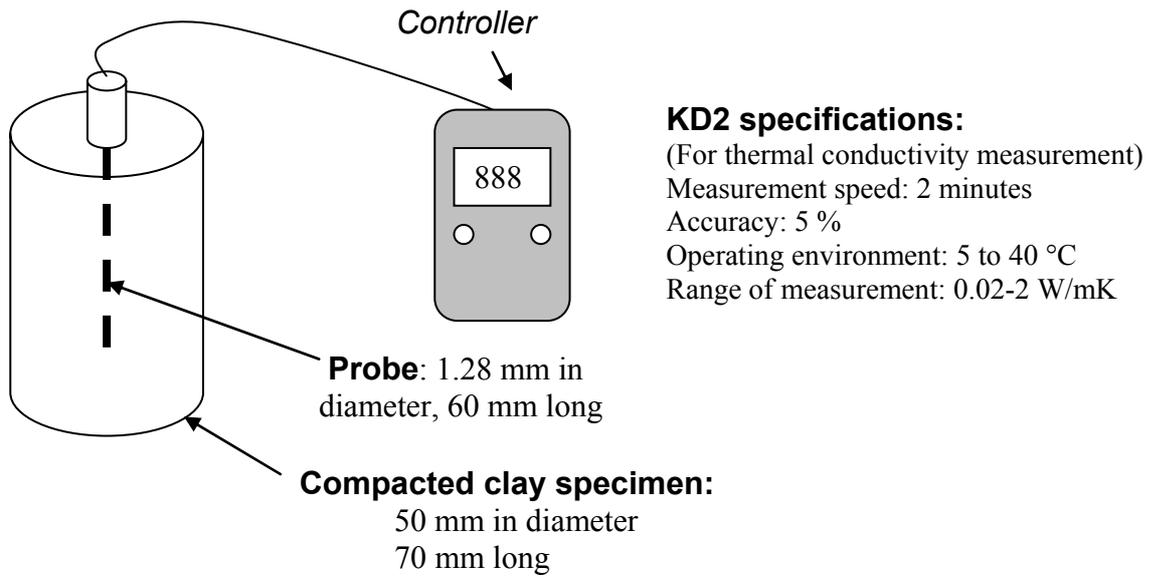

**Figure 1. Measurement of thermal conductivity.**

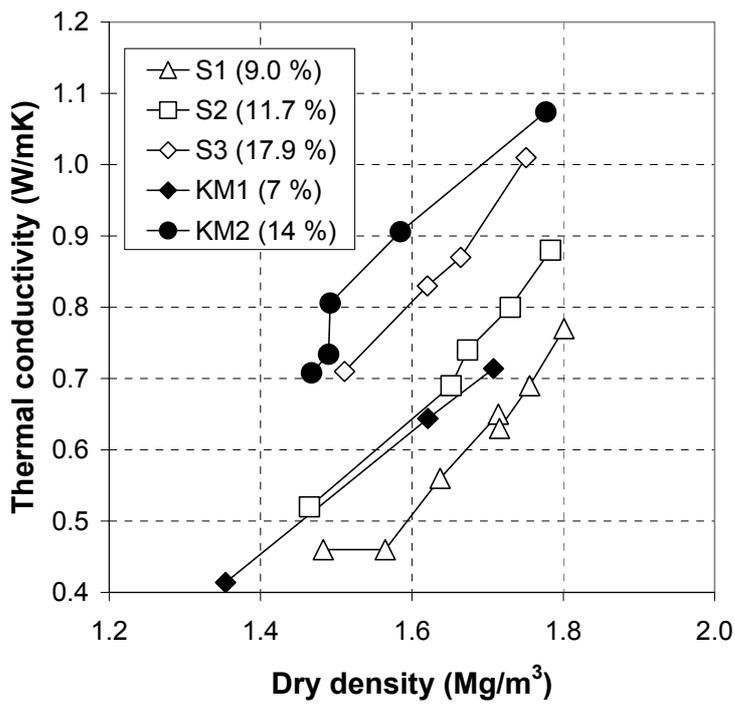

**Figure 2. Thermal conductivity versus dry density.**

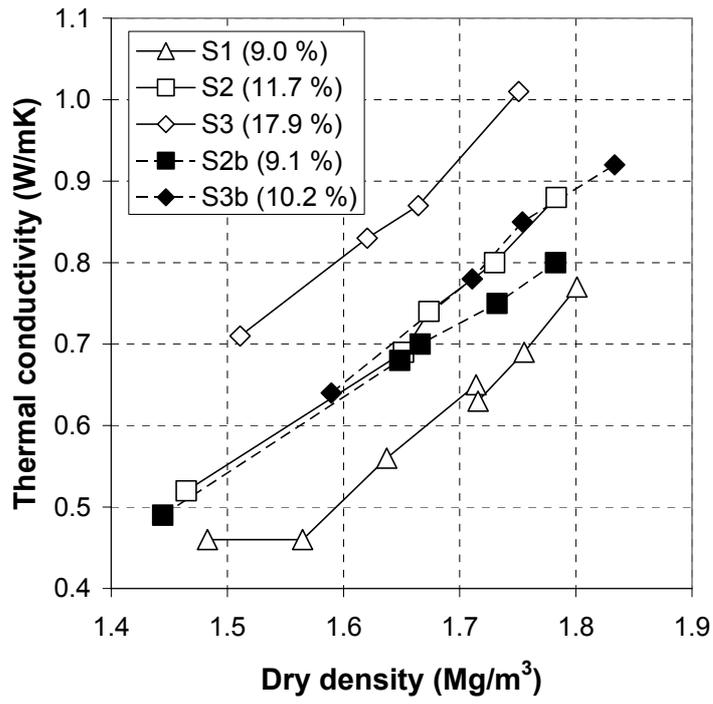

**Figure 3. Effect of drying on the thermal conductivity – dry density relationship.**

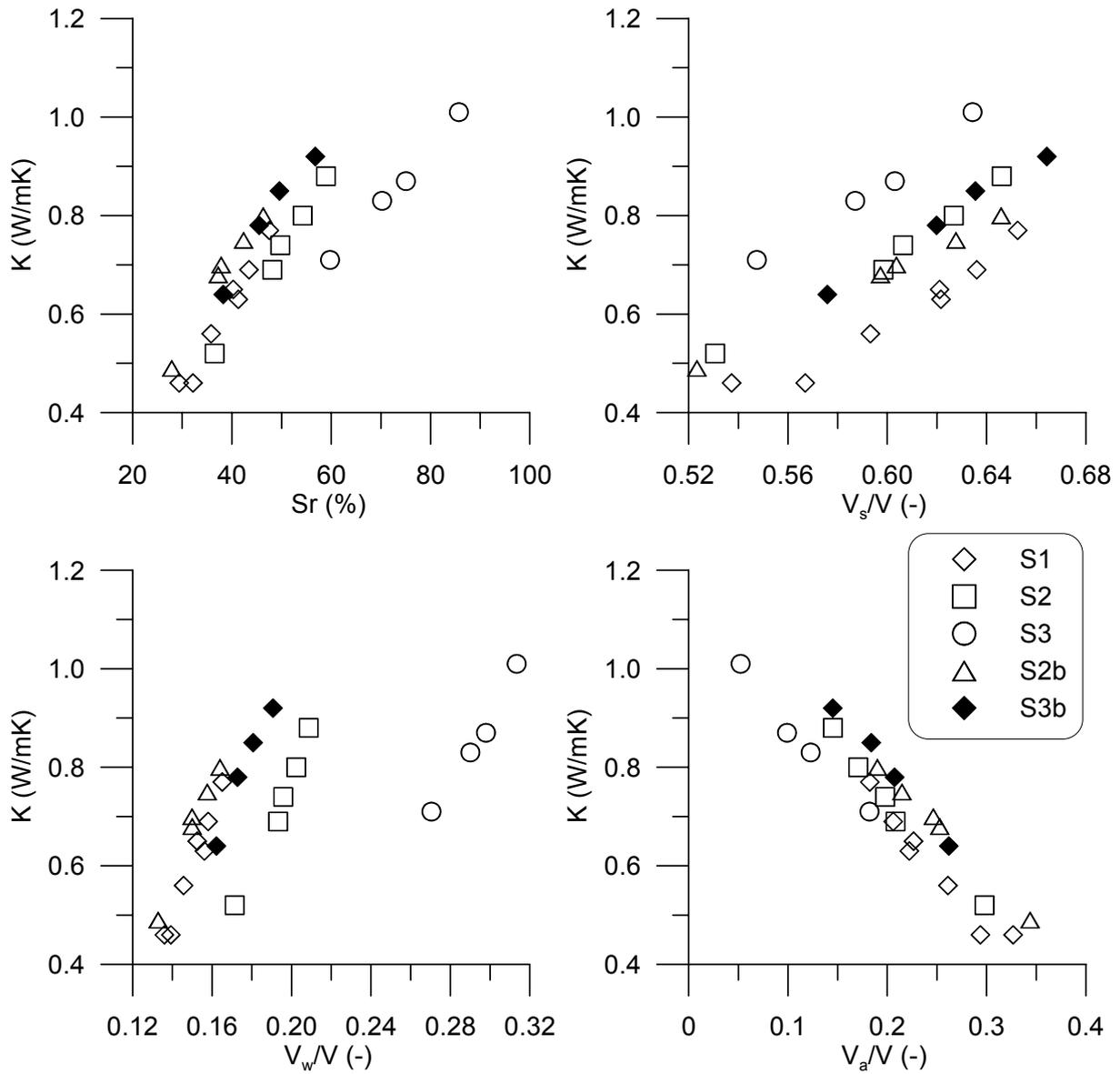

**Figure 4. Thermal conductivity versus (a) degree of saturation ; (b) volume fraction of solid; (c) volume fraction of water; (d) volume fraction of air.**

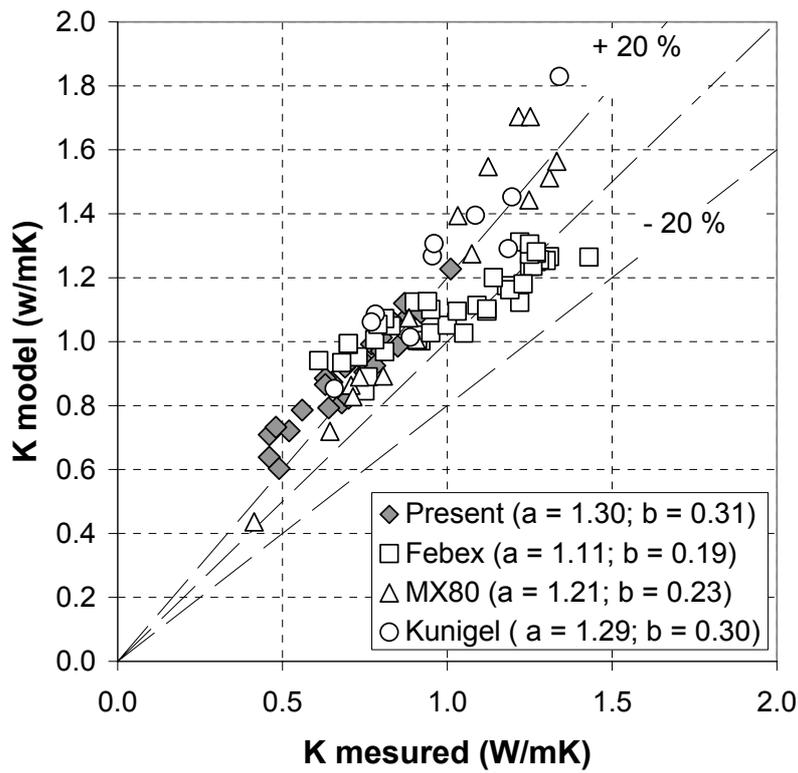

**Figure 5.** Thermal conductivity calculated by Johansen's method versus measured values.

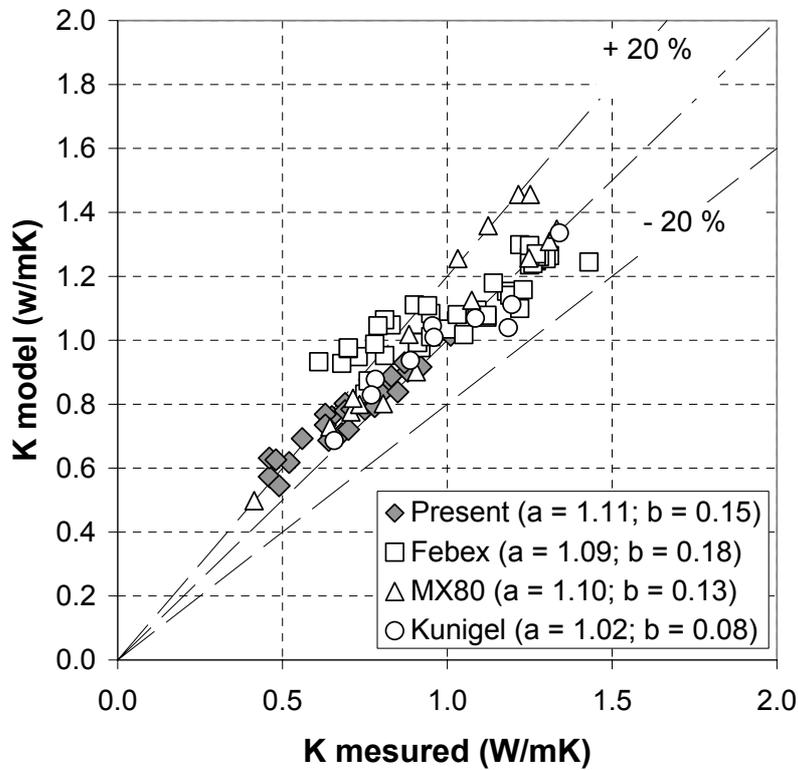

**Figure 6.** Thermal conductivity calculated by De Vries's method versus measured values.

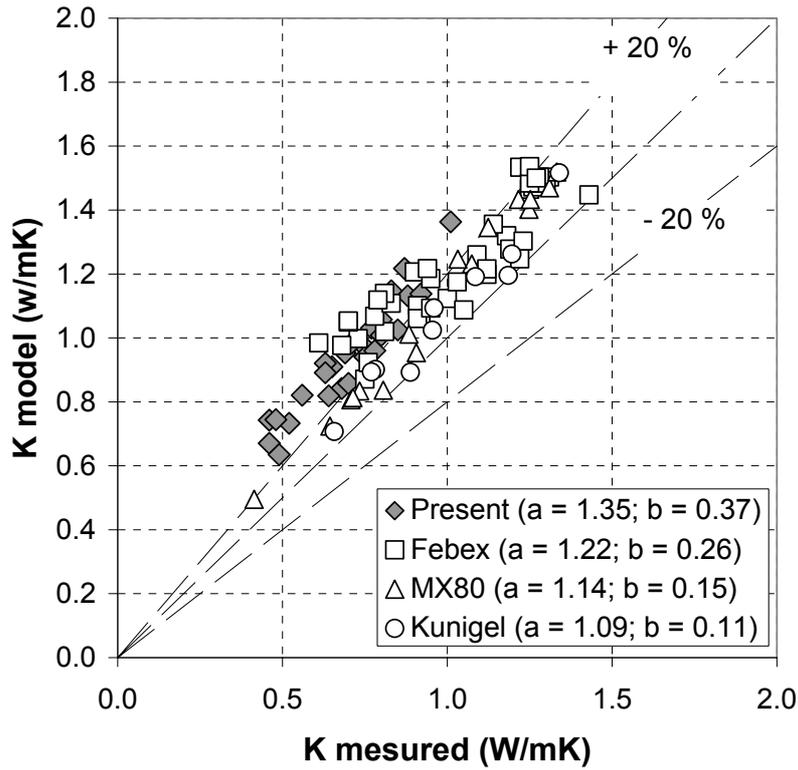

**Figure 7. Thermal conductivity calculated by Sakashita and Kumada's method versus measured values.**

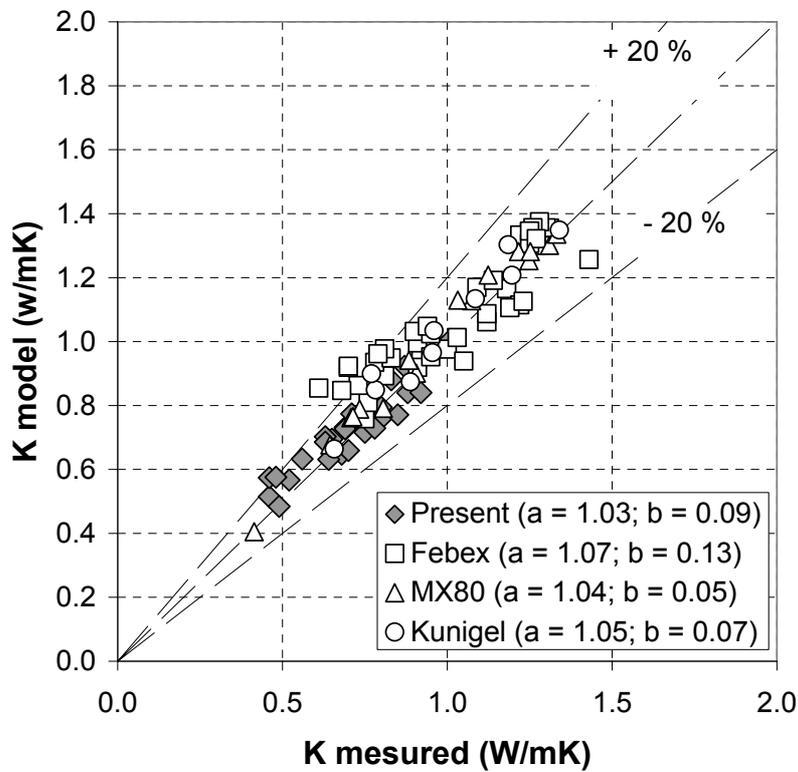

**Figure 8. Thermal conductivity calculated by the linear correlation versus measured values.**